# The intriguing mechanism of phosphorus anodes for sodium ion batteries revealed by *operando* pair distribution function and X-ray diffraction computed tomography

Jonas Sottmann[†], Marco Di Michiel[‡], Helmer Fjellvåg[†], Lorenzo Malavasi[#], Serena Margadonna[§*], Ponniah Vajeeston[†], Gavin Vaughan[‡], David S. Wragg[†*]

[†] Department of Chemistry, University of Oslo, Blindern, P.O. Box 1033, 0315 Oslo, Norway.
[‡] ESRF - The European Synchrotron, CS40220, 38043 Grenoble, France.
[§] College of Engineering, Swansea University, Swansea SA1 8EN, UK.
[#] Department of Chemistry, University of Pavia and INSTM, Viale Taramelli 16, 27100 Pavia – Italy.

*corresponding authors

*Note that after the first author, the author names are listed in alphabetical order. A full list of author contributions is given at the end of the paper.*

**Abstract**

Phosphorus is one of the most promising anodes for sodium ion batteries (SIBs). Little is known about the structural mechanism of Na/P cycling due to the fact that only one of the structures involved ($Na_3P$) is crystalline. Using *operando* X-ray diffraction computed tomography (XRD-CT) analysed by Rietveld and pair distribution (PDF) methods combined with density functional theory (DFT) calculations we show that the sodiation and desodiation mechanisms of phosphorus are very different. Sodiation follows the thermodynamic path of lowest energy from P via NaP to $Na_3P$ while desodiation follows a kinetically controlled deintercalation mechanism in which the layered $Na_{3-x}P$ type structure is maintained until P nanoclusters form. Using XRD-CT allows analysis of the 3D structure of the anode, but most importantly, removes the contributions from the sample container and other battery components, particularly important for PDF analysis.

**Introduction**

If we could store the energy produced from renewable sources effectively then the demand for fossil fuel energy generation would be drastically reduced. Sodium ion batteries (SIBs) are one of the best technological hopes for stationary electrochemical energy storage, offering a cheap and chemically similar alternative to lithium ion batteries (LIBs). The key weakness in adapting LIB technology for SIBs is the anode material. Graphite is a highly effective anode for LIBs but has a poor capacity for Na: new materials are needed. Among the possible solutions alloying anodes have proved one of the most promising, with high capacities based on the fact that each anode atom can fix multiple Na ions. Several metallic and semi-metallic elements have been tested as SIB anodes, but the highest capacity has been exhibited by red phosphorus (theoretical capacity 2596 mAh $g^{-1}$). This system (reported almost simultaneously by two groups) furthermore shows a reversible capacity of up to 1890 mAh $g^{-1}$ as a composite with carbon and considerable cycling stability with a capacity of around 1000 mAh $g^{-1}$ retained after 140 charge/discharge cycles[1-2]. Li et al showed that moderate cycling stability can also be obtained by forming a composite of red phosphorus with carbon nanotubes[3]. More recently Sun and co-workers have reported graphene/phosphorene hybrids with a specific capacity of 2440 mAh $g^{-1}$ (calculated using the mass of phosphorus only) at a current density of 0.05 A $g^{-1}$ and an 83% capacity retention after 100 cycles[4]. They found that the phosphorene layer is transformed to an amorphous structure during the first cycle while still being confined between graphene layers. Xu et al reported a composite of black phosphorus, "Ketjenblack" carbon polymer and multi-walled carbon nanotubes and investigated the mechanism of sodiation and desodiation using *ex situ* NMR and TEM, revealing NaP as an amorphous intermediate phase[5]. Ramireddy and co-workers made a systematic study of the existing data on P/C composite anodes for LIBs and SIBs and added their own measurements including TEM[6].

The mechanisms of bismuth and antimony alloying anodes have recently been elucidated using combinations of X-ray diffraction, total X-ray scattering, spectroscopy and calculations[7-8]. The mechanism of phosphorus anodes, however, remains a subject of debate- in large part due to the fact that of all the possible phases present during battery cycling, only $Na_3P$ is observed in crystalline form. The study on Sb anodes by Allan and co-workers points the way here- using pair distribution function analysis of *operando* battery total scattering data they elucidate the different phases found during the transformation from Sb to $Na_3Sb$ and back again, including non-crystalline, non-stoichiometric intermediates[7]. Recently some of the same authors also used calculation methods to determine the most stable phases expected during the cycling of phosphorus anodes in LIBs and SIBs[9].

One of the problems in studying battery structures in general and in particular when using total X-ray scattering methods is the contribution to the data from parts of the battery other than those in which we are interested. Allan et al[7] and Chapman et al[10-12] have addressed this by using background subtraction methods and adapting their batteries significantly for total scattering experiments. Here we propose an alternative and potentially much more informative method based on the PDF-CT approach first reported by Jacques et al[13]. Although operando tomographic imaging has recently revealed significant information on the macrostructure of batteries[14-16], the technique has not previously been applied using atomic structural data. Jensen et al have studied commercial lithium ion and nickel metal hydride batteries using XRD-CT methods, mapping the distribution of $LiCoO_2$ from the intensity of a single peak and obtaining information on its orientation from the 2D diffraction patterns[17], but not extracting actual structural data. By reconstructing the entire battery system from Rietveld and PDF analysis of XRD-CT data we have, firstly, removed all contributions from the sample container, electrolyte etc. and secondly, obtained three dimensional spatially resolved atomic structural data on a working phosphorus anode- revealing with full clarity the different mechanisms of sodiation and desodiation of phosphorus in good agreement with cycling data. The technique described here can easily be combined with traditional X-ray tomographic imaging (see ESI) to provide additional information on the battery macrostructure.

**Methods**

Amorphous phosphorus was prepared by milling red phosphorus (99.99 %, Sigma Aldrich) in a Fritsch Planetary Micro Mill Pulverisette 7 at 720 rpm with a ball-to-powder ratio of 20:1 for 24 h. A composite of amorphous phosphorus with CNTs (purified and multi-walled, n-tec) was formed by milling them in a mass ratio of 7:3 in a Fritsch Mini-Mill Pulverisette 23 at 50 Hz with a ball-to-powder ratio of 10:1 for 20 min. The working electrode was prepared by spreading slurry composed of 70 wt % of the composite, 10 wt % of conductive carbon black (Super P, Timcal) and 20 wt % poly(acrylicacid) (PAA, Sigma Aldrich) as binder dissolved in ethanol on the Al pistons used in the *operando* sample container. Drying of the electrodes was carried out at 60 °C overnight. The electrodes were thereafter handled under inert atmosphere. The working electrode was separated from the Na metal disk as counter electrode by electrolyte soaked glass fibres (GF/C, Whatman). As electrolyte a 1 M solution of $NaPF_6$ in ethylene carbonate/diethyl carbonate (EC/DEC, 1:1 in wt) solution with the addition of 5 wt % FEC was prepared. All electrolyte constituents were purchased from Sigma Aldrich. The battery was galvanostatically cycled in a voltage range of 0.01 V to 2 V vs $Na/Na^+$ using a Biologic SP150 with low current option. The specific capacity values are expressed on the basis of the mass of phosphorus.

Half cells were assembled in a specially constructed X-ray transparent electrochemical container for *operando* XRD/PDF-CT and absorption tomography measurements (see ESI). The sample container consists of a sealed Teflon cylinder containing two Al pistons on which working and Na metal counter electrode are directly deposited. The electrodes are separated by glass fibres soaked with electrolyte. The sample container is aligned such that the working electrode layer (containing about 0.1 mg of amorphous phosphorous and being about 30 μm thick) is oriented in plane with the X-ray beam and can be rotated by up to 360° around its vertical axis during the measurement.

Data were collected on beamline ID15A of the European Synchrotron (ESRF). The XRD/PDF-CT data were collected at an energy of 69.8 KeV (λ = 0.1779 Å) with a beam size of $16 \times 200 \mu m^2$ (V×H). Absorption tomography data were collected with an energy of 46.3 KeV, field of view $1.3 \times 1.3 mm^2$ and pixel size of $1.2 \times 1.2 \mu m^2$ ). XRD/PDF-CT tomograms were collected with 5 vertical slices spaced by 8 μm followed by an absorption tomography measurement. The raw 2D diffraction images were azimuthally integrated to give 1D powder diffraction patterns which were used in the tomographic reconstruction. The strategy used for collection and reconstruction of the diffraction tomographic slices is described elsewhere[13, 18-19]. The data collection took about 7 min per slice for XRD/PDF-CT, while each absorption tomography measurement took about 15 min. The sequence of measurements was repeated several times during the first desodation and second sodiation of the Na/P half-cell, which lasted about 4 h.

Corrections were applied for artefacts in the filtered back projection tomogram caused by the Teflon walls of the sample container. A three dimensional map of the battery absorption coefficient was measured by a conventional absorption micro-tomography scan. An XRD-CT scan of the empty sample container was then measured. The absorption caused by the sample on the Teflon signal of the empty sample container was calculated using the 3-D absorption map. The calculation was done using a ray tracing program which computes the absorption correction for the diffracted signal generated by each voxel as a function of the sample orientation, 2-theta angle and azimuthal angle. The absorption corrected Teflon signal was removed from the battery diffraction signal for each voxel. The corrected diffractograms were processed using PDFGETX3[20] to give radial pair distribution functions. Further details of the reconstruction and corrections are given in ESI section S2.

Rietveld refinements were carried out using a "surface" strategy (i.e. treating all the diffractograms as a single 3D "surface" of data with some parameters linked for the entire dataset but no use of parametric equations) based on the parametric Rietveld method in TOPAS V5[21-22]. The model of Brauer and Zintl for $Na_3P$[23] retrieved from the ICSD was used as the starting point. Background, lattice parameters, atom positions, peak broadening (Lorentzian crystallite size based on a fundamental parameters peak shape) thermal parameters and scale factors were refined simultaneously for all diffraction patterns in each group (usually a complete slice of the tomogram[24]) while the zero error and tan-θ broadening[25] were refined as single parameters for the whole dataset. Adsorption was handled using the parallel beam capillary correction in TOPAS with a fixed adsorption value. A typical fit is shown in the ESI (figure S16) along with the Rwp values for the fits (table S2).

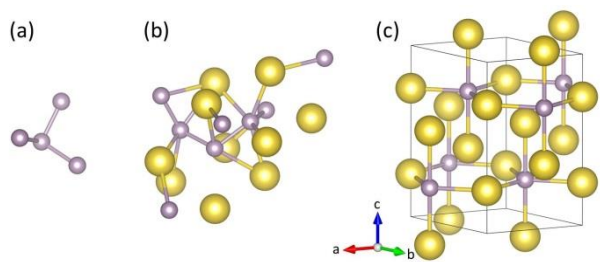

Figure 1 Models used in the PDF analysis: (a) amorphous P nano-cluster, (b) amorphous NaP nano-cluster and (c) crystalline Na3P.

The PDF data were analysed using routine adapted from diffpy-CMI[26]. Based on the observations from the Rietveld analysis that there was limited variation within the slices and only slices 2 and 3 were composed mainly of anode material, PDFs from the relevant voxels (i.e. those in which sample material was observed) were averaged for each of these slices. The scattering data were rebinned with a linearly increasing bin size in order to reduce the effects of statistical and systematic noise; this can be done without loss of information as the patterns vary smoothly and are devoid of sharp features at high Q. The step size was varied such that the final step at high Q was five times larger than the step at low Q. Background due to cell and sample container constituents was determined by using the residuals from the most featureless patterns (those containing only amorphous P), after subtraction of the electrode signal. The resulting background G(r) was then fitted to a spline in order to avoid biasing the subtraction to noise, and subtracted from the G(r) of all of the signals. The crystallographic model for $Na_3P$ was the same as that used in the Rietveld refinements, while the NaP and phosphorus cluster models were built up from the models of Hönle and Von Schnering (NaP)[27] and Hultgren et al (red phosphorus- one of the first example of PDF analysis)[28]. A spherical particle approximation was used to account for small domain size in the $Na_3P$ refinements, and the size of the clusters of NaP and P were estimated from the extent of the PDFs at the relevant stages of reaction and the clusters were then edited by trial and error using the Vesta package[29] and tested until the best fits were found. The final cluster models used are shown in Figure 1. The scale factors were refined for all phases and lattice parameters were refined for the $Na_3P$, the only phase with significant long range order. A global bond length variation parameter was refined for the NaP and P clusters. In order to reduce the number of free parameters, thermal parameters for the Na and P were constrained to be constant during all fits and across all of the phases. The instrumental parameter "qdamp" was held constant during all refinements, as was "delta2" for each phase.

Total energies were calculated by the projected-augmented plane-wave (PAW) implementation of the Vienna *ab initio* simulation package (VASP).[30-31] The methods used for the calculations are described more fully in ESI section S3.

**Results and discussion**

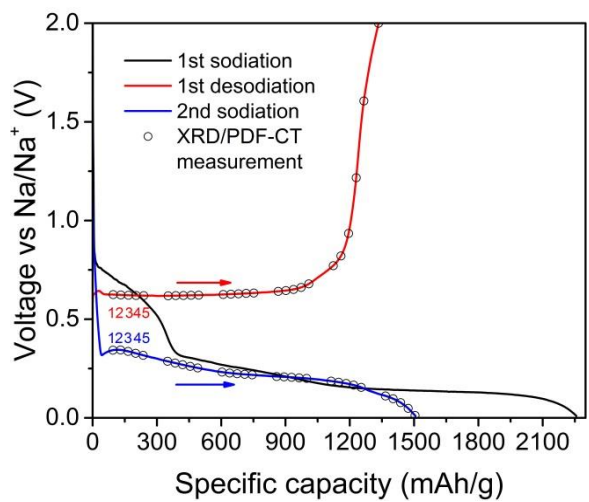

Figure 2. Voltage profile for cycling of the Na/P half-cell during the *operando* experiment. Circles indicate the voltage/capacity states for the measured XRD/PDF-CT slices. Each XRD/PDF tomogram comprises a set of 5 slices.

The voltage profile of the Na/P half-cell during the experiment (first sodiation, desodiation and second sodiation) is shown in Figure 2. Note that the first sodiation of the cell, during which the irreversible formation of the solid electrolyte interface (SEI) occurs was not studied in the *operando* tomographic experiment.

The desodiation profile has a single long plateau at about 0.6 V. During the second sodiation voltage plateaus are observed at about 0.3 and 0.2 V, as for the latter stages of the first sodiation. These features suggest different sodiation and desodiation mechanisms. Qian et al observed several voltage plateaus during desodiation for P/C composites, corresponding to well defined cyclic voltammetry (CV) peaks, which they attributed to a series of phases ($Na_3P$ - $Na_2P$ - $NaP$ - $NaP_7$) following the Na/P phase diagram[32]. It is possible that these extra desodiation plateaus were not observed by ourselves or other researchers due to the very low rate used by Qian et al for their CV measurement (0.02 mV/s ≈ C/30)[1]. They assume sodiation to follow the same path, although the voltage plateaus/CV peaks are less well defined. Kim et al[2] present voltage profiles similar to ours (although their desodiation plateau slopes between around 0.4 and 0.6 V) as do Li et al[3] and Ramireddy at al[6]. These authors do not speculate on the mechanisms beyond pointing out that the plateau at 0.2 V during sodiation corresponds to the formation of crystalline $Na_3P$, observed in *ex situ* XRD patterns. The phosphorene/graphene sandwich anode reported by Sun et al shows a small plateau at ~0.7 V during desodiation in addition to the long 0.6 V plateau[4]. In fact this plateau is present to some degree in all Na/P cycling data but is often hard to see (note the slight change in slope at a capacity of around 1000 mAh $g^{-1}$ in Figure 2). Their sodiation voltage profile (after the first sodiation) is very similar to that of our sample. The voltage profile observed by Xu et al[5] is consistent with ours, although the small plateau they attribute to NaP occurs at slightly higher voltage than the corresponding feature in our data.

Reconstruction of the sample using X-ray absorption tomography methods allowed us to locate the areas of interest in the cell, while also revealing significant movements in the cell components during cycling, due in some part to the expansion of the phosphorus anode (a movie depicting this is included in the ESI).

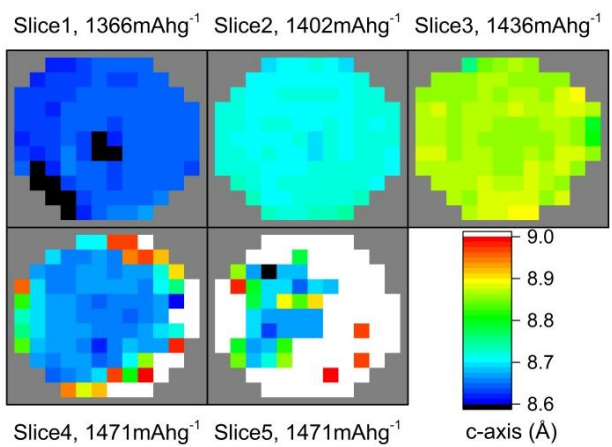

Figure 3 Tomographic slices at the end of the second sodiation, reconstructed on the *c*-axis of Na$_3$P extracted by parametric Rietveld methods. Slices were collected a different distances from the current collector. Slice 1 is closest to the current collector and slices 4 and 5 partly cut into the glass fibre separator. Note that the variation within the slices is minimal. The *c*-axis variations between the slices are probably due in the most part to changes in the charge state during the 7 min acquisition time per slice.

Initial analysis of the reconstructed data using "surface" Rietveld methods (with the crystalline Na$_3$P structure) showed relatively small structural variations within the individual tomographic slices (Figure 3), but confirmed the stability of the methods and the location of the anode material at the sodiated stages of cycling. The results show clearly that slices 4 and 5 of the tomogram contain significant amounts of glass fibre separating material, with slice 5 containing virtually no anode material. Variations in the parameters between the layers are probably due principally to the changes in charge state during the 7 minute acquisition time per slice (see the charge state labels in Figure 3). The central section of slice 4, however, does seem to have a well determined Na$_3$P *c*-axis parameter which is similar to that of slice 1 (close to the current collector). This may indicate that the reaction is at a different stage at the outer slices of the anode, close to the electrolyte and current collector, than in the central slices of the anode. Note however that the sodiation is complete at slice 4, with the final capacity of 1471 mAhg-1 reached, while in slices 1, 2 and 3 sodiation is still in progress. This is supported by comparing the reconstructed diffractograms from the centre of the layers (ESI, figure S17), which appear to be of similar quality for slices 2, 3 and 4. It is possible that the beam size used (16x200μm$^2$) in this experiment relative to the small crystallite sizes (see below) results in some averaging of the structures- it may be feasible to resolve structural variations on a smaller scale using a smaller beam.

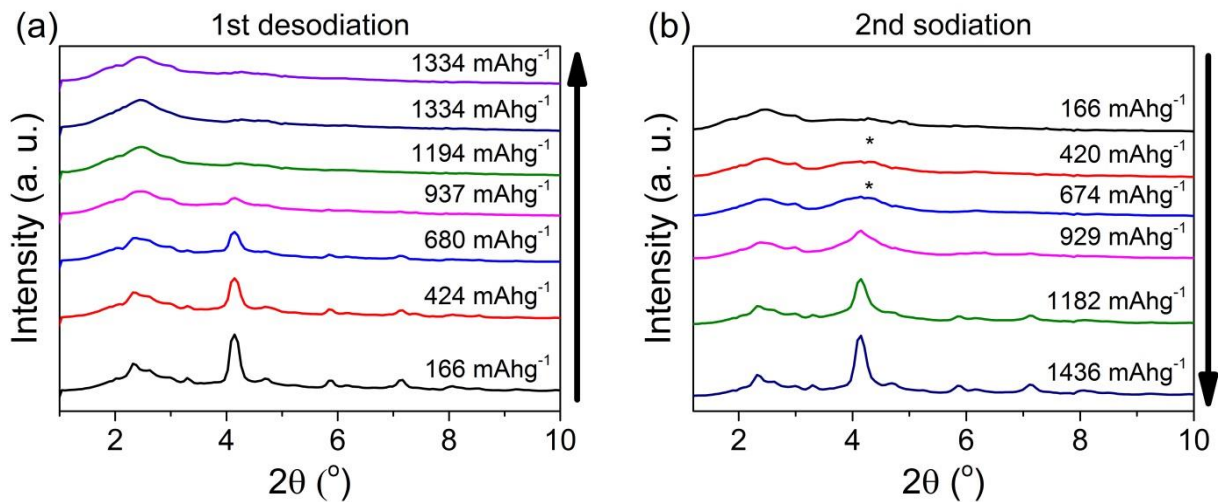

Figure 4. Reconstructed XRD data of voxel at position (15, 15) in tomographic slice 3 at different charge states for the first desodiation (a) and second sodiation (b). The broad feature observed at 2θ ≈ 4 ° is marked with *.

As variation within each slice was minimal, further analysis focussed on the central voxels of the slices at different stages of reaction. Reconstructed diffractograms for desodiation and second sodiation are shown in Figure 4. The Bragg peaks from the Na$_3$P phase are clearly visible when the anode is fully sodiated, but it is also important to note the broad feature in the diffraction patterns at 2θ ≈ 4 ° which appears during sodiation but not during desodiation (we will show below that this almost certainly represents NaP). The broad peak at ~2.5 ° is due to phosphorus nanoparticles.

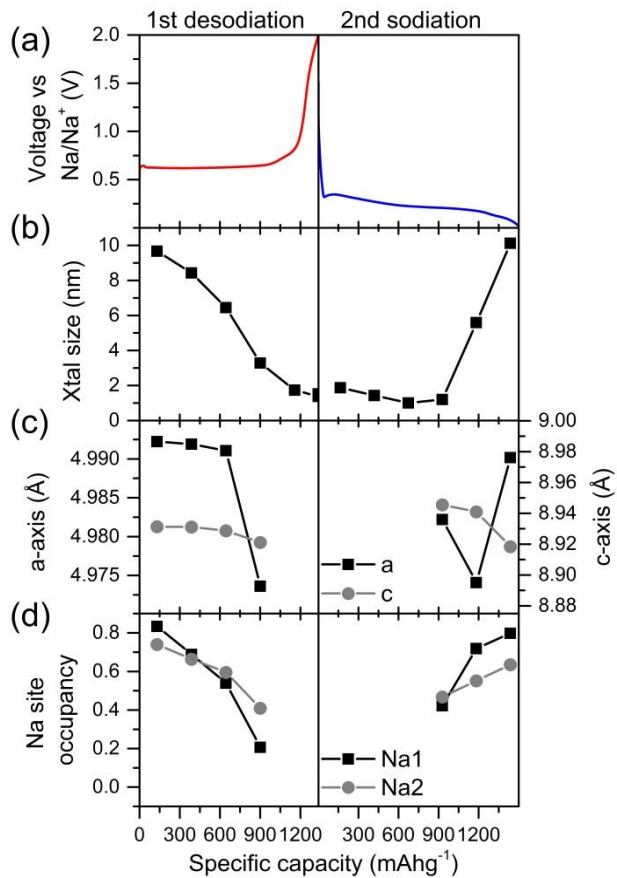

Figure 5. Voltage profiles for desodiation and sodiation (a) and data from Rietveld refinements of Na$_3$P in voxel at position (15,15) of tomographic slice 3: crystallite size (b), lattice parameters (c) and Na site occupancy (d). Lattice parameter and site occupancy results from diffractograms collected at charge states where crystalline Na$_3$P is not present are omitted.

Rietveld analysis of the voxel at position (15,15) in tomographic slices 2 and 3 showed the appearance of the highly crystalline Na$_3$P phase during sodiation and its disappearance during desodiation, best observed from the Lorentzian crystallite size broadening parameter. Trends in the crystallite size, unit cell axes and occupancy of the Na atom sites of the layered Na$_3$P structure are shown in Figure 5 along with the voltage profiles for comparison. A clear decline in crystallite size is observed during desodiation with a corresponding growth in the crystallite size during sodiation at the point where the phase begins to appear. Crystallite sizes in nm were obtained with the double-Voigt approach using the LVol-IB method from a fundamental parameters peak shape in TOPAS version 5. (Note that although the Lorentzian crystallite size macro was used in the TOPAS input files to model the peak broadening, although it is quite possible that the true source of broadening is strain). We were not able to fully characterise the setup for fundamental parameters crystallite size determination, and the size values from the Rietveld analysis are therefore not absolute; the trends however are real. The two Na sites are seen to depopulate during desodiation and gradually repopulate as the Na$_3$P phase reappears during sodiation. This suggests high mobility of Na in the phase, in agreement with the findings of Qian et al[1]. Similar high Na mobility was observed by Allan et al in Na$_3$Sb[7]. In addition we see that the occupancies of the Na sites never reach 100%, suggesting that fully stoichiometric Na$_3$P is not formed. The lattice parameter data is somewhat surprising, in that during the first desodiation *a* and *c* both decrease slightly as the sodium level is reduced, while during the second sodiation *c* declines with sodiation while the *a*-axis shrinks, then extends. The results for slice 2 are very similar (ESI, figure S18).

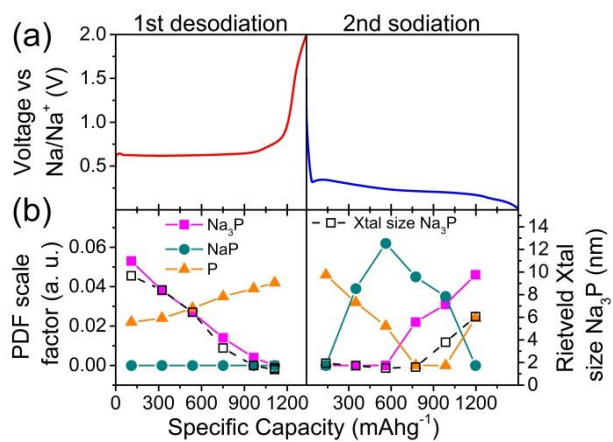

Figure 6. Voltage profiles for desodiation and sodiation (a) and plots of the PDF scale factor for all phases (b). Crystallite size obtained from Rietveld refinement of Na$_3$P is plotted for comparison.

The fitted PDFs (G(r)) of the reconstructed diffractograms are shown in the ESI (Figure S3-15), along with the fit agreement factors (table S1). From the extent of the peaks in the PDF it is possible to extract crystallite/cluster sizes at each charge state. Refinements showed that the data could be fitted using linear combinations of the PDFs for crystalline Na$_3$P and nano-clusters of NaP and P. Scale factors for each phase are plotted in Figure 6 with the voltage curves for desodiation and second sodiation above. The Lorentzian crystallite size (used to indicate the presence of crystalline Na$_3$P as the Rietveld scale factor is highly influenced by increases in the peak width when there are no longer clear Bragg peaks in the patterns and therefore fails to clearly indicate the appearance and disappearance of the crystalline phase) from the Rietveld refinement is plotted on the same axes to allow comparison between the PDF and Rietveld analysis for Na$_3$P. The G(r) when fully desodiated has only one peak at approx. 2.25 Å (the approximate length of a P-P bond) indicating a phosphorus cluster of just a few atoms with no order beyond the first shell of P-P bonds. We fitted this using a model cluster of just 4 P atoms (see Figure 1).

The PDF analysis reveals that NaP is formed during sodiation but not during desodiation (meaning that our assignment of the broad peak at 2θ ≈ 4 ° in the powder diffraction patterns in Figure 4 to NaP is correct), while Na$_3$P is present for longer during desodiation. The crystallite size of Na$_3$P from the Rietveld refinements was used as an indicator of whether the phase was observed in the Bragg scattering. Compared to the Rietveld analysis of Na$_3$P we observe the phase later during desodiation and earlier during sodiation in the PDF data, suggesting that amorphous Na$_3$P exists in addition to the amorphous NaP and P phases.

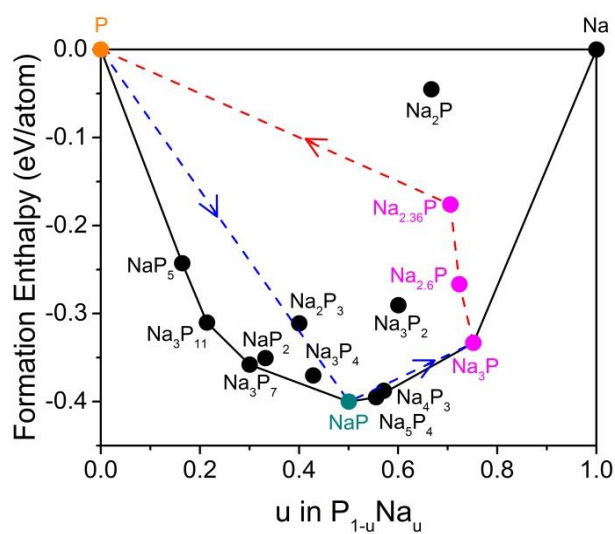

Figure 7. DFT convex energy hull diagram showing the lowest energy phases in the P-Na phase diagram and formation enthalpies for Na deficient Na$_{3-x}$P structures (coloured in magenta). The black line indicates the lowest energy route, while the blue and red dotted lines show the observed routes of desodiation and sodiation, respectively.

DFT calculations were used to construct a convex energy hull for the Na/P system and this was found to be in agreement with that reported by Mayo et al.[9] The hull suggests the most thermodynamically favourable route between P and Na$_3$P. In agreement with our PDF results, the lowest energy route to Na$_3$P passes through the rather stable NaP helical P-chain structure (Figure 1b). This acts as an intermediate in a two phase type mechanism, easing the transition through to layered Na$_3$P. We have so far found no evidence for the presence of any of the other stable Na/P phases lying on or close to the energy hull at any stage of cycling, though this cannot be entirely ruled out. It is possible that they may be observed with better time/space resolution and that the tiny clusters may exhibit varying stoichiometry. The Na$_5$P$_4$ phase (derived from the Na$_5$As$_4$ crystal structure[33]), with chains of 4

P-atoms, seems to be a likely candidate (during sodiation), lying on the Na/P convex hull. These phases would ease the transition from very small P clusters to the layered structure of $Na_3P$ (see ESI). The recent NMR work of Xu et al also shows clear evidence of NaP as an intermediate structure[5].

The energy hull does not help us to understand why the thermodynamic route is not followed on desodiation. We therefore carried out further calculations on the deintercalation of Na from $Na_3P$, maintaining the layered structure as observed in the *operando* tomographic experiment. We found that several stable phases exist, suggesting that a smooth transition from $Na_3P$ through $Na_{3-x}P$ to P should be favoured by kinetics over the nucleation of NaP. i.e. in the case of high Na mobility as suggested by Qian et al[1] and our Rietveld refinements, NaP should not form. $Na_{2.6}P$ and $Na_{2.36}P$ are found to be the possible stable Na deficient structures in the $Na_3P$ matrix and are included as points in the energy diagram (the red dotted line indicates the kinetically favoured path from $Na_3P$ to P). Furthermore, we calculated that $Na_2P$, which might serve as an intermediate on the path to NaP during desodiation, is highly unstable (this phase is also shown, well above the convex hull, in Figure 7). We believe that the small plateau observed at ~0.7 V during desodiation for phosphorene/graphene anodes[4] is a sign that the thermodynamic mechanism ($Na_3P$-NaP-P) co-exists with the kinetic deintercalation mechanism to some degree at lower rates of desodiation. A similar kinetic mechanism was observed during desodiation of the Na/Sb system by Allan and co-workers[7].

The reasons for the different mechanisms of sodiation and desodiation of the group 15 elements are still unclear, though size effects may be significant. P (0.44 Å) and Sb (0.90 Å) have sodiation routes different from those of their desodiation, which include significant amorphous components. The somewhat larger Bi (1.03 Å[34]) follows the same route on sodiation and desodiation and retains crystallinity at all stages, with the crystal structure of $Na_3Bi$ depending on the crystallite size[8]. We also note that the size of Bi is similar to that of $Na^+$ (1.02 Å effective ionic radius), while P and Sb are smaller. No experimental data on the Na/As system is available, although gallium arsenide has recently been studied as an anode for LIBs[35]. The rate of deintercalation of Na from the crystalline $Na_3P$ is probably too rapid for the formation of intermediates except at very low rates of desodiation[1,4]. $Na_3Sb$ is similar, with the deintercalation leading to other clearly identifiable amorphous intermediates, but not NaSb[7]. The $^{23}Na$ NMR results reported by Allan and co-workers show high Na mobility (which they relate to the excellent high rate performance of Sb anodes) in crystalline $Na_3Sb$ which supports the idea that fast Na deintercalation may be connected to the differences between sodiation and desodiation routes in P and Sb anodes[7]. The thermodynamic sodiation route from P nanoclusters to NaP chains to layered $Na_3P$, involves the breaking and forming of many bonds and should be significantly slower than deintercalation. A similar scheme with kinetically and thermodynamically controlled routes for charge and discharge respectively is described by Kim and co-workers for the $Na/Co_3O_4$ battery system[36].

**Conclusions**

By using a combination of *operando* XRD and PDF computed tomography on a working half-cell in tandem with DFT calculations, we have clearly revealed the different structural mechanisms of sodiation and desodiation of a phosphorus anode. Sodiation takes place via a thermodynamically favourable two-phase type mechanism with an amorphous, chain type NaP intermediate; while desodiation proceeds via rapid, kinetically controlled deintercalation of $Na^+$ from $Na_3P$, in which the crystalline layered structure of $Na_3P$ is preserved while amorphous P is formed until a significant amount of the Na$^+$ is removed. Indeed, the Na$_3$P structure is observed in the PDF analysis after its crystallinity is lost. It is likely that this occurs in (generally successful) competition with the reversal of the thermodynamically driven sodiation mechanism.

The reaction path followed by the material is probably determined by the size of the alloying anode metal, the mobility of Na$^+$ in the fully sodiated form (observed in the variation of Na occupancy during the *operando* experiment) and the stability of Na$_3$P, which allow the desodiation to follow a kinetically rather than thermodynamically favourable path.

The tomographic method allowed us to focus purely on the anode material to obtain the best possible data for PDF and Rietveld analysis and build up a 3-D picture of the anode based on *operando* atomic structure information extracted by both methods. We believe this technique can offer very significant benefits for future structural studies of crystalline and amorphous battery materials.

**Author Contributions**

SM and LM conceived and prepared the experiment; JS, MDM and SM performed the measurements; JS did the sample preparation and electrochemical characterisation; JS and MDM designed the sample container, processed and corrected the tomographic data; DSW performed the Rietveld refinements; GV performed the PDF analysis; PV performed the DFT calculations; the manuscript was prepared by DSW, with additions from the other co-authors.